\pdfoutput=1

\documentclass[11pt]{article}

\usepackage[]{acl}

\usepackage{times}
\usepackage{latexsym}
\usepackage{pgfplots}

\usepackage[T1]{fontenc}

\usepackage[utf8]{inputenc}

\usepackage{microtype}

\usepackage{graphicx}
\usepackage{multicol}
\usepackage{multirow}
\title{FinTech for Social Good: \\ A Research Agenda from NLP Perspective}

\author{Chung-Chi Chen,\textsuperscript{1} Hiroya Takamura,\textsuperscript{1} Hsin-Hsi Chen\textsuperscript{2}\\
 \textsuperscript{1} AIST, Japan \\
 \textsuperscript{2}  Department of Computer Science and Information Engineering \\ 
 National Taiwan University, Taiwan \\
  \texttt{c.c.chen@acm.org, takamura.hiroya@aist.go.jp,} \\ \texttt{hhchen@ntu.edu.tw}\\
  }

\begin{document}
\maketitle
\begin{abstract}
Making our research results positively impact on society and environment is one of the goals our community has been pursuing recently.
Although financial technology (FinTech) is one of the popular application fields, we notice that there is no discussion on how NLP can help in FinTech for the social good. 
When mentioning FinTech for social good, people are talking about financial inclusion and green finance. 
However, the role of NLP in these directions only gets limited discussions. 
To fill this gap, this paper shares our idea of how we can use NLP in FinTech for social good. 
We hope readers can rethink the relationship between finance and NLP based on our sharing, and further join us in improving the financial literacy of individual investors and improving the supports for impact investment.
\end{abstract}

\section{Introduction}
\label{sec:introduction}

Recently, our community is eager for the discussions on AI for social good\footnote{\url{https://ijcai-22.org/calls-ai-for-good/}} and NLP for social good.\footnote{\url{https://2021.aclweb.org/calls/papers/}} 
Although we have several surveys and opinion pieces from different aspects~\cite{hovy2016social,leins2020give,jin2021good}, there is hardly any discussion on how NLP can help in FinTech for social good development.
As the FinTech industry matures, we think that it is time to discuss the potential risk and what we can do to address these issues with the help of NLP.

When mentioning FinTech for social good, there are many discussions on how FinTech development can improve financial inclusion~\cite{ozili2021financial}. 
Financial inclusion is a notion to express that everyone can get basic financial services and have an opportunity to access financial products. 
With the development of mobile banking and wireless internet, more and more individuals in developing countries can access financial services such as cashless payment, deposits, and insurance~\cite{analytics2018financial}.  
With the development of online identity authentication, people in developed countries can easily access financial products such as equity and options by opening an account within a few minutes. 
Both are important achievements in the FinTech industry. 

However, we notice some attendant risks following the improvement of financial inclusion. 
Based on the statistics of the Taiwan Stock Exchange (TWSE), people in their twenties occupy over 80\% of accounts opened. 
On the other hand, based on the statistics of the Taiwan Academy of Banking and Finance\footnote{\url{https://www.tabf.org.tw/English/}}, a government-supervised institution, approximately 40\% of people in their twenties are financially illiterate. 
Even those who graduated from a university, only 5\% of them are considered to have a good sense to finance. 
Although joining the financial market could make people become wealthy, it may also lead to bankrupting results. 
Given 80/20 Rule, i.e., 80\% of individual investors will only get profit and loss balance outcome or deficit, which is supported in financial market~\cite{xiao2015does}, can we do something to protect individual investors?
One of the recent examples\footnote{\url{https://edition.cnn.com/2021/02/11/investing/robinhood-lawsuit-suicide-alex-kearns/index.html}} also indicates that making trading process becomes easy and game-like may lead to negative outcomes when the user does not have enough knowledge to understand the risk of investment and the results of these actions. 
Following this line of thought, we want to discuss how we can improve financial literacy with FinTech and NLP (FinNLP).

Taking the environment and society into consideration when making daily decisions is one of the tendencies for sustainable development. 
It is the same in financial decision-making. 
Instead of only considering return and risk, some people in the financial domain advocate adding social and environmental impact to the decision-making process. 
To be the bridge between the financial field and the FinNLP community, we also make discussions on how FinNLP can support the development of impact investment.

In the rest of this paper, we will first discuss current FinNLP research and the role of these studies in FinTech development in Section~\ref{sec:Current FinNLP Research}. 
In Section~\ref{sec:FinNLP for Social Good}, we propose two directions, investor education and impact investment, for FinNLP for social good. 
Some challenges and possible solutions will be provided in Section~\ref{sec:Challenges and Directions}. 
Finally, we conclude this paper in Section~\ref{sec:conclusion}. 
Our intention is to propose some different directions, which got few discussions before, to inspire researchers to figure out some ways for FinNLP for social good together.

\section{Current FinNLP Research}
\label{sec:Current FinNLP Research}
FinTech is an emerging term from 2015, and the development of FinNLP has three major directions: improving the working processes in financial institutions, predicting market information, and providing automatic financial services to customers~\cite{chen2020nlp}. 
Improving the working processes can enlarge the capability of financial institutions to serve more people, and providing automatic financial services can improve the customers' experience. 
Both are the key forces in achieving financial inclusion. 

From the aspect of improving working processes,  one of the widely discussed topics is identity fraud detection because it is a crucial issue when people use services remotely. 
One ideal solution is to ask some new questions that could not be answered based on user-provided data. 
\citet{wang2019you} share the experimental results of this idea, and generate questions based on extended personal knowledge graphs to improve performance. 

Forecasting market information is the most popular direction in FinNLP development. 
Stock movement~\cite{xu-cohen-2018-stock,tang2021retrieving}, volatility~\cite{rekabsaz-etal-2017-volatility,dereli-saraclar-2019-convolutional,qin-yang-2019-say,10.1145/3459637.3482089}, sales~\cite{lin-etal-2019-learning}, bubble~\cite{sawhney-etal-2022-cryptocurrency}, and several different kinds of market information are targets of market information forecasting tasks. 
In addition to constructing end-to-end models for prediction, there are some other discussions on how to fool NLP-based market prediction models~\cite{xie-etal-2022-word} and the problems of pre-trained language models in market information forecasting applications~\cite{chuang-yang-2022-buy}.

Instead of giving predictions directly, leveraging NLP techniques to  provide customers information for them to make the final decision is a human-centered approach. 
Some systems are proposed for this purpose. 
\citet{liou2021finsense} propose a system to help investors and journalists identify the related companies based on a given news article. 
\citet{hassanzadeh2022knowledge} present a comprehensive system for news retrieval, event identification, causal analysis, and causal knowledge graph extraction.  
Users can make decisions based on the outputs of these systems.

\section{FinNLP for Social Good}
\label{sec:FinNLP for Social Good}
Thanks to the effort of researchers in the FinNLP community,  models can better understand financial documents than before, and some application scenarios can be implemented for the customers of financial institutions. 
However, how to make social good based on our findings has been hardly discussed so far. 
In this section, we propose two directions: investor education and impact investment. 
Investor education is a way to improve financial literacy, which is one of the world-level projects proposed by the Organisation for Economic Co-operation and Development (OECD)\footnote{\url{https://www.oecd.org/finance/financial-education/}}.
Impact investment is a concept that adds one more dimension ---impact on the environment and society--- into consideration when making investment decisions. 
The goal of these two directions is to construct a sustainable financial environment. 
In this section, we provide the link between these topics and FinNLP.

\subsection{Investor Education}
\label{sec:Investor Education}
Currently, investor education is provided passively. 
For example, the United States Securities and Exchange Commission (SEC) provides many teaching materials for investors,\footnote{\url{https://www.sec.gov/education/investor-education}} and TWSE also makes many efforts to prepare materials for investor education\footnote{\url{https://investoredu.twse.com.tw/Pages/TWSE.aspx}}.
We think that NLP provides an opportunity to change from passive to active. 
Instead of asking investors to take a look at the materials, we can score their investment ideas and provide feedback on their ideas. 
Thus, for investor education, we propose two directions: scoring investment reasons, and generating risk reminders.

Scoring opinions is not a new topic in our community. 
For example, scoring persuasive essays~\cite{ghosh-etal-2016-coarse} and evaluating the helpfulness of product reviews~\cite{ocampo-diaz-ng-2018-modeling} have been discussed for a long time. 
However, scoring investment opinions is still in the early stage. 
\citet{ying-duboue-2019-rationale} provide a pilot exploration on scoring investment reasons by annotating 2,622 rationales into four levels. 
\citet{chen2021evaluating} evaluate amateurs' opinions based on whether they share some characteristics with experts' opinions. 
These studies also point out two different goals of scoring investment opinions: (1) scoring rationality and (2) using profitability as the score.

Since there are too many uncertainty factors in the financial market, it is hard to say whether the reasonable analysis will lead to profitable outcomes. 
Even professionals' analysis may sometimes be inaccurate~\cite{zong-etal-2020-measuring}. 
Therefore, the goal of scoring investment reasons is different from that of making predictions. 
We aim to inform investors whether their reasons can support their decision based on financial theories. 
Additionally, we can let investors know to what extent it will lead to a profitable outcome based on the learned results with historical data.

However, only providing the score is not enough for investors to learn from models' output. 
Providing feedback is more useful than just letting users know the score. 
For example, grammatical error correction~\cite{xie-etal-2018-noising} is one of the common next steps after scoring essays. 
Providing feedback to the grammatical error~\cite{nagata-2019-toward} is a more challenging task. 
In investor education, we think that providing risk reminders to the given investment reasons is more important than just providing scores.
For example, when users input their reasons as the follows: \footnote{Both opinion and risk reminders parts are selected from J.P. Morgan Asia Pacific Equity Research's report on 13 May 2022. }

\begin{quote}
    Key takeaways from Quanta’s 1Q22 analyst call included: 1) 1Q22 GMs dipped to an eight-quarter low due to inferior product mix and inefficient production; 2) weak 2Q22 earnings outlook with both PC demand weakness and supply constraints; 3) 2Q22 NB shipment guidance at 20\% qoq declines missed market expectations, implying likely m/s loss towards Hon Hai, in our view; 4) Server order book remains intact with double-digit growth target this year; 5) full-year capex at NT\$10bn with overseas expansion (Thailand, North America); and 6) the board approved a NT\$6.6 DPS, implying 8\% yields at current stock prices. Quanta stock has moved largely in line with Taiex YTD, while we expect the earnings downgrade (first time since 2019) to weigh on the stock price in the next six months. We cut 2022/23 earnings by 15\%/7\% and lower our Dec-22 PT to NT\$77. Stay \textbf{Neutral}.
\end{quote}

In addition to providing scores, we think that it will be more meaningful if models can generate the risk reminders to this neutral opinion as follows:
\begin{quote}
\textbf{Key downside risks} include margin erosion in servers
and a potential slowdown of PC demand post COVID-19.

\textbf{Key upside risks} include 1) a longer-than-expected server upcycle; and 2) stronger-
than-expected MacBook market share gains offset consumer PC weakness.    
\end{quote}
This kind of risk reminder is different from the outputs of end-to-end market movement prediction models. 
It provides some scenarios that may make the neutral claim become inaccurate. 
We think that it is an ideal way to provide feedback on financial opinions. 
In this way, investors can take the generated risk into consideration when making decisions. 
It will also help investors consider their investment reasons more comprehensively.

\subsection{Impact Investment}
Beyond considering return and risk in the investment, the impact on the environment and society has been widely discussed and suggested to add to the financial decision-making process~\cite{emerson2012risk,chiappini2017social}. 
More and more concrete ideas for assessment and auditing are proposed. 
Recently, one of the most popular guidelines is SASB Standards, proposed by Sustainability Accounting Standards Board.\footnote{\url{https://www.sasb.org/}}
This standard sorts out the non-monetary issues related to environmental, social, and governance (ESG) that may influence the financial performance of companies. 
It is intuitive that NLP can help in capturing ESG-related terms~\cite{kang-etal-2021-finsim} or tracking semantic change in companies' reports~\cite{purver-etal-2022-tracking}.
However, can FinNLP help more on impact investment? 
We propose two directions for this question: non-monetary opportunity/risk identification and scenario planning.

As we shown in Section~\ref{sec:Current FinNLP Research}, many studies pay attention to monetary return or volatility prediction. 
It can be based on news articles, social media, professional reports, meeting between investors and company managers, and so on. 
However, limited studies try to identify the opportunity and risk from non-monetary aspects, i.e., ESG aspect. 
Since the impact period of ESG-related events may be longer than that of traditional monetary events, such as the increase of earning, decrease of sales, etc, we think that one of the important directions is to identify whether the given event is the opportunity or risk to ESG and how it will influence companies' sustainable operation. 
As shown in \citet{10.1145/3511808.3557585}, simple statistical methods perform better than neural network models in multivariate long sequence time-series forecasting. 
That raises an open issue: whether their claim holds true when we attempt to evaluate long-term non-monetary impacts? 
Since many market information forecasting studies mainly pay attention to short-term prediction (e.g., next day, 3-day, and 1 week), we want to encourage more studies to pay attention to long-term value assessment.

In addition to identifying the opportunity and risk, we think that scenario planning (analysis) is one of the key directions that FinNLP can help in impact investment. 
The goal of scenario planning is to understand the possible impact of the event better rather than to predict the future. 
For example, in addition to identifying whether it is a risk to the agricultural industry when a piece of news related to ``climate anomalies'' is given, we expect models to generate some plausible scenarios such as ``it would lead to lack of water'', and this scenario can be extended to ``it may influence the agricultural industry''.

Besides the most possible scenario, the worst-case scenario should also be generated for discussion. 
That is, when performing scenario planning, experts first generate many scenarios regardless of the probability of occurrence. 
And then, experts will select a few key scenarios for discussions and try to figure out some solutions to these key scenarios. 
Because scenario planning plays an important role in both monetary and non-monetary based financial decision, we think that it is a good direction to explore for providing better help to both individual and industrial investors.

\section{Challenges and Opportunities}
\label{sec:Challenges and Directions}

Because we just start in the proposed directions, there are many open research questions and opportunities.
In this section, we highlight three specific topics. 

\subsection{Evaluation}
One of the conveniences of market information forecasting tasks is that we can use market information as ground truth. 
It is open access and can be quickly obtained. 
However, when exploring the proposed risk reminder and scenario planning tasks, we should not just focus on the final outcome in the market. 
For example, the key downside/upside risks shown in Section~\ref{sec:Investor Education} may not happen finally, but these reminders are important for leading investors to have careful consideration before making decisions. 
In other words, previous studies are precision-focused studies, but the proposed directions aim to highlight the importance of covering as more as crucial factors as possible. 
Human evaluation is an intuitive direction.
However, how to automatically evaluate the generated scenarios is still an open issue.

\subsection{Implementation}
It is hard to imagine that an analyst says, ``I have read a lot of news from 2008 to 2010, analysis reports from 2003 to 2012, cooperate filing from 1994 to 2019, and earnings call transcripts from 2004 to 2019. Based on these experiences and yesterday's Made by Google product launch, I think Pixel Watch would be the hottest product in 2022.''
However, we are now using pretrained FinBERTs~\cite{araci2019finbert,huang2022finbert} in this way. 
There is no doubt that some common senses and word meanings can be learned via pretrained processes. 
However, the financial market changes all the time, and there is a bulk of information daily. 
How to add the latest knowledge and news into models becomes a challenge when we want to score investment reasons and generate risk reminders. 
One of the possible ways is using the knowledge in textual book and formal material to construct knowledge graph (KG) as the base of background knowledge.  
And then, we can further link this KG with the temporal-aware KG~\cite{goel2020diachronic}, which tracks the change of latest information. 

\subsection{Utilization}
Currently, the FinTech industry provides more convenient interfaces to investors, and also makes financial information become transparent. 
Many trading bots have been constructed to displace individual investors when making trading decisions. 
In this paper, we propose a human-centered AI concept with FinNLP, which aims to empower human abilities instead of replacing human. 
We think that the proposed directions for investor education are the next step for improving the investment environment of individual investors. 
There are many open research questions when exploring the proposed directions. 
To what extent state-of-the-art market information prediction models can help in scoring investment reasons? 
What is the difference between general storytelling tasks and financial scenario generation? 
We also think that investor education and current FinNLP studies are reciprocal. 
For example, scoring rationales can help us filter out low-quality opinions for opinion-based market information prediction models. 
It would be helpful in improving the performance of existing tasks.

\section{Conclusion}
\label{sec:conclusion}
In this paper, we point out the risk behind the recent improvement in financial inclusion, and share a research agenda from FinNLP for social good aspect. 
We hope this paper can lead readers to rethink the role of NLP in the financial field. 
Different from recent FinNLP studies, which focus on (1) extracting cooperation reports, analyst reports, news articles, and so on, and (2) using the extracted information for constructing economics/financial indicators or end-to-end market information prediction models, we propose two directions that could help improve financial literacy and augment investors' ability in making impact-oriented financial decisions. 
We are not saying the proposed directions are completely new, but we want to guide our community to consider to what extent recent findings can be applied to these directions and what we can do to achieve these goals. 
Our intent is that this paper can set an example for others to follow, and future works can further propose more ideas and solutions for FinNLP for social good.

\bibliography{custom}
\bibliographystyle{acl_natbib}



\end{document}